\documentstyle[preprint,aps]{revtex}
\begin{document}
\draft
\title{The Photon-Box Bohr-Einstein Debate Demythologized}
\author{A. C. de la Torre, A. Daleo and I. Garc\'{\i}a-Mata
%\footnote{dltorre@mdp.edu.ar}
}
\address{Departamento de F\'{\i}sica,
 Universidad Nacional de Mar del Plata\\
 Funes 3350, 7600 Mar del Plata, Argentina\\
dltorre@mdp.edu.ar
}
\date{\today}
\maketitle
\begin{abstract}
The legendary discussion between Einstein and Bohr concerning  the
photon box experiment is critically analyzed. It is shown that
Einstein's  argument is flawed and Bohr's reply is wrong.
%\par \noindent
%KEY WORDS: .
\end{abstract}
%\pacs{PACS 03.65.Bz  03.65.Ca}
\narrowtext
\par
\noindent
I. INTRODUCTION
\par
The disagreement between Bohr and Einstein concerning quantum
mechanics has become legendary in physics. One of their
discussions, perhaps the one that has received most attention, is
about the famous photon box experiment, a {\em Gedankenexperiment}
devised by Einstein in order to show an apparent flaw in quantum
mechanics: a violation of the time-energy indeterminacy relation.
Bohr's reply to it, constructed during a sleepless night after
Einstein's presentation, used the red shift formula, a result of
general relativity. This discussion has been publicized as a
crucial moment in the Bohr-Einstein  debate. Bohr himself said
that  the ``discussion took quite a dramatic turn''\cite{bohr}.
Although some authors have criticized the validity of Bohr's
reply\cite{jam} there is a general belief that the photon box
experiment was the arena of a master fight between two giants
(this view was also held by us).  However, in this work we will
see that this reputation is highly undeserved. Indeed a careful
and irreverent study of the issue shows that Einstein's argument
is flawed and Bohr's reply is wrong. There are some indications
that neither Einstein nor Bohr were satisfied by their part in the
discussion. Most probably Einstein noticed that Bohr's reply was
not conclusive; however he did not insist with his argument,
probably because he did no longer believe in it. On the other
hand, it is reported\cite{bohr1} that Bohr had, on the day of his
death, a drawing of the photon box in his blackboard. This could
be thought to be a hunter's trophy but it is also possible to
interpret this as a sign that Bohr was not satisfied with his
reply and was still looking for something better. The
Bohr-Einstein discussion  concerning the photon box has been
treated by many authors, either criticizing or justifying and
improving Bohr's argumentation. To our knowledge, none of them
makes a critical discussion of Einstein's argument. This is
probably due to the fact that his  argument is extremely simple
and apparently it requires no further comment. We will see however
that his argument is seriously flawed, but not for the reasons
given by Bohr.
\par
Although the core of the discussion took place at the Solvay meeting
in Brussels in 1930, we will take
as an authorized source for the argument and the reply,
Bohr's account of them presented in his article published
in 1949 in a book that has become a standard reference\cite{bohr}.
Einstein read this article and had the
opportunity (not used) to comment on these issues in the same
volume\cite{eins}.
\par
In this work we will use the terms ``indeterminacy'' and
``uncertainty'' with different meaning. The first term,
``indeterminacy'', denotes the impossibility of assigning precise
values to the observables of a system as prescribed by quantum
mechanics; whereas  ``uncertainty'' will refer to the lack of
precision in the knowledge of the value assigned to an observable
due to apparatus or experimental limitations. According to this
convention, Heisenberg's relations refer to indeterminacies and
not to uncertainties. Therefore, ``uncertainty'' will have in this
work a classical origin  and its nature is gnoseological whereas
``indeterminacy'' is essentially quantum-mechanical, regardless of
whether its nature is ontological or gnoseological, an issue not
decided  among the experts in the foundations of quantum
mechanics. In particular, referring to the indeterminacies either
in space-time or in energy-momentum, two opposite points of view
can been taken. In one of them it is claimed that the particles do
have a precise localization in energy and momentum as well as in
space and time, but quantum mechanics is unable to calculate or
predict them simultaneously. This position implies that quantum
mechanics is an incomplete theory. In the opposing view, quantum
mechanics is a complete theory but the particles do not have the
classical property of having exact values assigned to all
observables.
\par
\noindent
II. EINSTEIN'S ARGUMENT
\par
 Einstein proposed to consider a box with perfectly reflecting
walls containing electromagnetic radiation. Inside the box, an
ideal clock mechanism could open a shutter at a predetermined
moment $T$ for a time $\Delta T$ short enough to let only one
photon escape. Therefore since $\Delta T\rightarrow 0$, the time
of emission of the photon can be, according to Einstein, exactly
known. Before and after the emission, we could weigh the box at
all leisure with unlimited precision and from the difference in
weight we could exactly determine the energy $E$ of the photon.
Therefore, argued Einstein, the escaping electromagnetic radiation
violated the relation
\begin{equation}
\Delta E \Delta T \geq \hbar \ ,
\end{equation}
where $\Delta E$ is the indeterminacy in the energy of the photon
and $\Delta T$ is the indeterminacy in the moment of emission. The
meaning of the time energy relation has been a subject of profound
analysis. The main difficulty with it is due to the fact that
``time'' is not a quantum mechanical observable and relation (1)
can not be derived from a commutation relation as is done with
position-momentum indeterminacy relations. For the purpose of this
work we don't need to worry about these difficulties because there
are ways to derive time-energy indeterminacy relations that do not
require the existence of a time operator\cite{tamtam}. In any
case, the violation of this relation, with the meaning  given
above would be fatal to quantum mechanics.
\par
If Einstein's argument were right, it would  not only be fatal for
the indeterminacy principle in quantum mechanics but it would {\em
also}  present an unsolvable contradiction between Einstein's  own
concept of a photon and  classical electrodynamics!  Indeed, if
the shutter is open during a vanishing time interval (for just
{\em one photon} to escape, Einstein thought) then the
electromagnetic pulse must be very sharp, ideally a Dirac delta.
According to classical electrodynamics, the Fourier components of
such a pulse involve a wide spectrum of frequencies. Therefore the
electromagnetic pulse does not have a precisely defined frequency.
On the other hand the unique escaping photon should have,
according to Einstein, a precisely defined energy, that is, a
precisely defined frequency ($E=h \nu$) in contradiction with the
sharp pulse. At least in 1949, Einstein was well aware  of this
contradiction as he stated that ``...indivisible point-like
localized quanta of the energy $h\nu$ (and momentum $h\nu/c$)...
contradicts Maxwell's theory''\cite{eins2}. We know today that the
photon concept is  compatible with Maxwell's theory provided that
we abandon the simultaneous requirement of point-like localization
and precise energy-momentum.
\par
In order to appreciate Einstein's argument and to understand its
weakness it is useful to review the history of the photon as a
quantum mechanical particle\cite{pais} and its relation to a
classical electromagnetic pulse. In the year 1905, when Einstein
postulated its existence, the photon was not considered to be a
{\em particle} but rather an indivisible ``parcel of
electromagnetic energy'' involved in the interaction with matter.
It was with the Compton effect, observed in 1923, that it became
clear that the photon, in its collision with an electron, was
exchanging energy {\em and} momentum, that is, individual particle
properties. Only in 1926, when the photon was more than 20 years
old, he was given his name\cite{lewis}. Today we may think of the
photon as a full fledged quantum mechanical particle, like an
electron or a neutrino,  obeying  typical indeterminacy relations.
A photon can be  prepared in a quantum state with precise energy
and momentum at the cost of loosing space-time localization. On
the other hand, a well localized photon escaping Einstein's box
through a shutter opened during a very short time, will have
unsharp energy. We should notice however that sometimes in the
bibliography the term ``photon'' has a more restrictive meaning,
being reserved for the cases of quantum states prepared with sharp
energy momentum (and therefore not localized). That is, those
states generated by creation operators applied to the vacuum
$a^{\dag}_{\mathbf{k}}|0\rangle$. With appropriate superposition
of these photon states of well defined energy, we can build states
corresponding to any desired configuration of the electromagnetic
fields. For instance, it can be shown\cite{electro} that the, so
called, coherent states correspond to a monochromatic linearly
polarized plane electromagnetic wave. The quantum mechanical
description of the electromagnetic field is a very rich subject
that we will not treat here. For our purpose it is sufficient to
mention that the space-time localization  of the quantum state
obtained as a  superposition of  photon states, is in
correspondence with the space-time width of a classical
electromagnetic pulse.
\par
From the proposed experiment, it follows that Einstein was
thinking of the photon as a sharply {\em localized} object with a
sharply defined energy (frequency). Such an object does not exist!
Einstein's experiment can not be realized, not even in {\em
gedanke,} because it involves a nonexistent  object. Of course one
can weigh the box before and after a short opening interval but
the results of these measurements can not be attributed to a
nonexistent physical system. Even if the ``localized photon'' did
exist, the measurement performed on the box would perturb its
state, as is today generally accepted, as consequence of the
projection postulate; no matter how far away the photon is.
Nonlocality is indeed the most astonishing aspect of the photon
box experiment.
\par
Einstein's argument would be applicable when the physical
system emitted by the box were a point like object with precisely
defined energy, like a classical particle for instance. But then,
the argument would be useless because it would not be surprising
that a classical apparatus with a classical physical system violates
quantum mechanics. Einstein {\em needed} the photon in his argument
because of its essential quantum nature. His error was to assign, to
this quantum mechanical object, classical features of localization
and sharp energy. This is not possible  as shown by the combination
of classical electrodynamics with the definition of the photon energy.
\par
\noindent
II. BOHR'S REPLY
\par
We have seen that the main weakness of Einstein's argument
lays in the ``photon'' side  because it requires a nonexistent
object. However Bohr looked for an error at another place:
at the box side.
The set of formulas used by Bohr in his reply to
Einstein in order to derive the inequality (1) is:
\begin{eqnarray}
\Delta E = & \Delta m\  c^{2} \ ,
\\
 \Delta p \ \Delta q \approx&\hbar \ ,
\\
\Delta p < & T \ g \ \Delta  m   \ ,
\\
\frac{\Delta T}{T} = & \frac{1}{c^{2}}\ g \ \Delta q \ .
\end{eqnarray}
This is a hybrid set involving classical mechanics (4),
special relativity (2), quantum mechanics (3) and (4) and general
relativity (5). We will see that this hybrid mixture is precisely
the root of the weakness of the argument.
Of course, with trivial manipulations of these formulas one can
readily arrive at the inequality (1). However, in order to provide
a proof of the inequality, the
relations (2) to (5) must be valid and the symbols used in these
formulas must have the same meaning as the one in the inequality (1).
We will see that these two requirements are not satisfied by Bohr's
reply.
\par
In Bohr's reply to Einstein, $T$ is the ``interval of balancing
procedure'', $\Delta m$ is the ``weighing ... accuracy'', $\Delta
q$ is the ``position ... accuracy'' and $\Delta p$ is the
``minimum latitude in the control of the momentum of the
box''\cite{bohr}. In these definitions there is a mixture of
classical uncertainties and quantum indeterminacies. We agree with
other authors that the symbol $\Delta$ in Bohr's arguments denotes
some unspecified measure of width whose nature (classical or
quantum) is not clearly stated. This ambiguity is typical of
Bohr's confusing argumentation style\cite{whita} that  was
tolerated because of his (deserved) authority. For this reason his
argument still ``raises many questions which have never been
satisfactorily answered''\cite{jan}. The main difficulty in this
task is that there is absolutely no unanimity as to the correct
reading of Bohr's argument. The number of different
interpretations of the argument and of the meaning of the several
symbols is astounding. In any case, the symbols involved in
relation (1) are quantum indeterminacies and whatever meaning Bohr
had in mind, sooner or later quantum indeterminacies must enter
the scene in order to end with relation (1). We will therefore
analyze Bohr's argument assuming that he also refers to quantum
indeterminacies. With this assumption we will show that Bohr's
argument is wrong.
\par
The first difficulty that we find with Bohr's argument is that the
symbol $T$ has not the same meaning in the set of relations (2) to
(5) as in relation (1). In Einstein's argument, $\Delta T$ is the
indeterminacy in the moment of escape of the ``photon'' (more
precisely, the time-width of the electromagnetic pulse) and in
Bohr it means the indeterminacy in the balancing time of the box
during the weighing procedure. These indeterminacies need not be
the same. The weighing of the box can, indeed, be made a long time
after the escape of the electromagnetic pulse. We have here
sufficient reason to take Bohr's reply as inconclusive.
\par
The next difficulty is that Bohr claims that the inequality (4) is
``obvious''. This  was not obvious to us and therefore we made a
bibliography search. Many authors simply quote Bohr literally
without further explanation. Other authors attempt some
explanations for it. In several cases an {\em equality} is derived
instead of the inequality (4), involving classical uncertainties
that are later replaced by quantum indeterminacies. None of the
authors consulted stated clearly the difference between classical
quantities causally related and essential quantum indeterminacies.
We reached the conclusion that relation (4) is {\em not obvious}
and, if valid, it must be proved with more care.
\par
Since relation (4) has been attributed to be a consequence of the
weighing procedure of the box in a spring balance, it is important
to analyze it in every detail.
 The procedure
consists in hanging  or taking away increasingly smaller weights
 until the box remains at rest with the pointer at the zero of the scale.
At this point the force of the spring cancels the total  weight of the
box and we can neglect them.
There is one subtle point worth mentioning about the weighing procedure.
In the light of the mass-energy relation, the weighing procedure
must be such that there can not be any energy dissipation to the
environment because this would imply a loss of mass. In particular, a
{\em damping} mechanism to stop the box could not be tolerated.
The weighing procedure must involve only  transfer of elastic,
gravitational and kinetic energy to stop the box without dissipation.
We assume that the required experimental skills exist for this to be
possible.
When the balancing weights get smaller, the movement of the pointer becomes
slower and longer balancing times are required.
The weighing terminates with an experimental precision $\delta m$ when
the addition or subtraction of such a mass would not produce any noticeable
displacement of the pointer during a previously chosen balancing time $T$.
If we are willing to spend more time we can reach better precision.  Clearly,
 the box can not be at absolute rest but it will be moving with
a  momentum $\delta p$ that must be negligibly small, so small
that during the time of balancing $T$, the displacement of  the
box $\delta q$ around the zero position should be smaller than any
perceivable distance. The gravitational force $g\delta m$
associated to the mass $\delta m$, acting during the time $T$
would cause then a change $\delta p$ in the momentum of the box.
That is,
\begin{equation}
\delta p = g\ \delta m \ T \ .
\end{equation}
We must emphasize the all quantities mentioned are classical
uncertainties typical of any measurement procedure. The mass
(energy) uncertainty $\delta m$ and the momentum uncertainty
$\delta p$ are causally related by the equation above and can be
estimated for a given experiment. They should not be confused with
quantum indeterminacies that are in general not causally related.
This equation is, formally, similar to Bohr's relation (4) and we
could be tempted to replace the uncertainties $\delta p$ and
$\delta m$ by the corresponding indeterminacies $\Delta p$ and
$\Delta m$ and somehow argue for the $<$ sign. We should resist
the temptation leading to a wrong result, as we will see. More
will be said below about this temptation. The closest that we can
get relating the classical uncertainties with the quantum
indeterminacies is to notice that the indeterminacies of the
quantum state of the macroscopic apparatus are of course much
smaller than the experimental uncertainties.
\begin{equation}
\Delta p < \delta p \ ;\ \Delta m < \delta m \ .
\end{equation}
Clearly, Bohr's relation (4) does not follow from relations (6)
and (7) above. Even more, we will next see that there are quantum
states of the box that violate relation (4).
\par
An appropriate {\em classical} model for a spring balance could be
a damped harmonic oscillator, if we can assume that the dissipated
energy (mass) is negligible compared to the weighed mass. For our
{\em quantum} case, however, the uncertainty in the measured mass
must be very small and we can not tolerate energy losses;
therefore we will take as a model for the box hanging from a
spring in a constant gravitational field a (undamped) harmonic
oscillator. The constant force of gravity is canceled by the
spring and can therefore be ignored (it just amounts to an offset
in the rest position of the box). The quantum state of the box
will be given then by the harmonic oscillator energy
eigenfunctions $\{\phi_{n}\}$. We may assume that the preparation
of the system, involved in the
 weighing procedure, leaves the box, including the enclosed radiation,
in the energy ground state $\phi_{0}$. This is the simplest
assumption but we can be more general and choose for the state of
the box some superposition of energy states with an energy
indeterminacy $\Delta E$ much smaller than the experimental
uncertainty $\delta m c^{2}$. Among these choices we could
consider the harmonic oscillator coherent states\cite{glau}
\begin{equation}
\psi_{\alpha} = \sum^{\infty}_{n=0} \exp(-\frac{|\alpha|^{2}}{2})
\frac{\alpha^{n}}{\sqrt{n!}}\ \phi_{n} \ .
\end{equation}
These states are particularly interesting because they are the
quantum states that most resemble the classical states in the
sense that they have the least indeterminacy product. Notice that
$\alpha$ can take any (complex) value and the ground state
corresponds to the special case where $\alpha=0$. The
indeterminacies in momentum and energy for these states are,
\begin{equation}
\Delta^{2}p = \hbar m \omega /2 \ ,\
\Delta^{2}E = \hbar^{2} \omega^{2} |\alpha|^{2} \ .
\end{equation}
Where $m$ is the mass of the oscillator (the box) and $\omega$ its
oscillation frequency. The momentum indeterminacy  is constant for
all states and taking $\alpha$ small enough we can reach a state
with an energy (mass) indeterminacy small enough to violate
inequality (4), providing a counter example to Bohr's ``obvious''
relation. That is,
\begin{equation}
|\alpha|< \frac{c^2}{T g}\sqrt{\frac{m}{2\hbar\omega}} \
\Longrightarrow \ \Delta p >  T \ g \ \Delta  m \ .
\end{equation}
Regardless of the meaning that Bohr gave to the symbols, we can
not use the set of relations (2) to (5), assuming that the
$\Delta$'s mean quantum indeterminacies, in order to derive
relation (1) because one of them is wrong, as the counterexample
shows. The last inequality above has the reversed sign compared
with relation (4) and, if the rest of Bohr's argument were
correct, it would lead to a result like (1) but with the reversed
sign! A boomerang for Bohr. With the harmonic oscillator model for
the box, a rather decent model, we have shown that relation (4) is
not generally true and presumably a model independent proof of the
relation is impossible. In any case, it is far from being
``obvious''.
\par
Here we would like to emphasize that quantum indeterminacies can not be
treated as classical variables because they are not causally related.
An indeterminacy in momentum $\Delta p $ {\em is not} a  change in
momentum that is caused by the action of some force. The indeterminacies
are just that, {\em indeterminacies} without a classical cause. For this
reason it would be wrong to place $\Delta m$ and $\Delta p$ instead of
$\delta m$ and $\delta p$  in Eq.(6). It follows from the formal
definition of the indeterminacies in quantum mechanics that when two
observables $A$ and $B$ are related by a function, $B=F(A)$, their
indeterminacies {\em are not related by the same function} (not even for a
linear function  with operator valued coefficients!). For example,
energy and momentum of a free particle  are related by $E=p^{2}/2m$ but
their indeterminacies are not related in the same way. As another example,
the position observables at different times
($t$ and $0$) for a free particle in a given quantum state, are related by
$x(t)=x(0)+vt$, where $v=p/m$
is the velocity observable. However in this case the indeterminacies in the
position at those times are related by
\begin{equation}
\Delta x(t) = \sqrt{ \Delta^{2}x(0) +\left(\langle xv+vx\rangle -
2\langle x\rangle\langle v\rangle \right) t+
\langle v^{2}\rangle \,t^{2} }\ ,
\end{equation}
where the expectation values are taken in the state at time $t=0$.
It is therefore wrong to use Heisenberg indeterminacy
relations in order to derive other relations by careless mathematical
manipulations of the indeterminacies. This illegal use of Heisenberg
relations has prompted D. Griffith  to say: ``when you hear a physicist
invoke the uncertainty principle, keep a hand on your wallet''\cite{grif}.
\par
\noindent
IV. CONCLUSION
\par
When we differentiate the meaning and the mathematical treatment of the
classical uncertainties and the quantum indeterminacies, it becomes
clear that it is impossible to prove a quantum mechanical relation,
like relation (1), by means of an argument concerning classical uncertainties.
If this were possible, then, quantum mechanics would be a consequence of
classical mechanics. On the other hand, we have just seen that hybrid
arguments mixing classical and quantum concepts are often meaningless
and may lead to wrong results.
Therefore the only way to prove the validity of a quantum relation
is to require logical consistency {\em within}  the quantum  theory
(and, of course, agreement with the experiments).
Every indeterminacy relation is a consequence of the formalism
of quantum mechanics where states are represented by Hilbert space
elements and observables by hermitian operators. Therefore any
correct quantum mechanical treatment of the photon box experiment
will be consistent
with the indeterminacy relations.
\par
Such  treatments are not the purpose of this work but we just want
to mention that they will involve the quantum description of the
electromagnetic radiation inside and outside the box with states
that are coupled by energy conservation. The description of the
outgoing radiation can be conveniently done in terms of states
built as superposition of photon number eigenstates\cite{electro}.
A well localized electromagnetic pulse will involve a large number
of photon states, implying a large energy spread.  On the other
hand, if energy is sharp, the electromagnetic pulse will not be
localized. Since the radiation inside and outside the box are
coupled by energy conservation, the state reduction involved in
the measurement of the energy inside the box will affect the state
of the radiation outside the box, no matter how far away it is.
Again, nonlocality appears as one of the most astonishing features
of quantum mechanics.
\par
In this work we have seen that the reputation  of the
Bohr-Einstein discussion concerning the photon-box experiment is
not justified. Their  arguments have been uncritically propagated
by many authors. It is unfortunate that these, very deficient,
arguments of Bohr and Einstein have been given such a high
priority in their debate. Indeed, arguments that are meant to
destroy quantum mechanics and the reply to save quantum mechanics
from destruction should have a level of rigor not reached in the
photon-box debate. Of course, the weakness in the arguments of
Bohr and Einstein are easily detected {\em today} in the light of
the present knowledge of quantum mechanics and it would be an
anachronistic error to blame them for that. These weakness should
by no means tarnish their fundamental contributions to quantum
theory. On Bohr's side, the idea of complementarity showing that
physical reality  has a level of sophistication veiled to our
naive observations; and on Einstein's side the EPR argument
exhibiting unexpected correlations whose study have dominated the
research on the foundations of quantum mechanics. Compared with
their gigantic contributions, their errors are insignificant. The
important mistake that we do want to point out is the uncritical
and authoritarian propagation of a coherent combination of errors
that lead to the right conclusion that quantum mechanics is
correct.
\par
This work has received partial support from ``Consejo Nacional de
Investigaciones Cient\'{\i}ficas y T\'ecnicas (CONICET), Argentina (PIP
grant Nr. 4342/96). A.D. and I.G.M. would like to thank the
``Comisi\'on de Investigaciones Cient\'{\i}ficas'' (CIC) for financial
support.

\end{document}